\documentclass{ecai} 

\usepackage{latexsym}
\usepackage{amssymb}
\usepackage{amsmath}
\usepackage{amsthm}
\usepackage{booktabs}
\usepackage{enumitem}
\usepackage{graphicx}
\usepackage{color}
\usepackage{subcaption}
\usepackage{makecell}
\usepackage{balance}
\usepackage{lastpage}
\usepackage{color, colortbl}

\newcommand{\BibTeX}{B\kern-.05em{\sc i\kern-.025em b}\kern-.08em\TeX}

\begin{document}


\begin{frontmatter}


\paperid{123} 



\title{AquaSignal: An Integrated Framework for Robust Underwater Acoustic Analysis}


\author[A]{\fnms{Eirini}~\snm{Panteli}\orcid{0009-0008-5747-8378}\thanks{Corresponding Author. Email: eirini.panteli@priorianalytica.com}}
\author[A]{\fnms{Paulo~E.}~\snm{Santos}\orcid{0000-0001-8484-0354}}
\author[A]{\fnms{Nabil}~\snm{Humphrey}\orcid{0000-0002-4227-2544}} 

\address[A]{PrioriAnalytica}


\begin{abstract}
This paper presents AquaSignal, a modular and scalable pipeline for preprocessing, denoising, classification, and novelty detection of underwater acoustic signals. Designed to operate effectively in noisy and dynamic marine environments, AquaSignal integrates state-of-the-art deep learning architectures to enhance the reliability and accuracy of acoustic signal analysis. The system is evaluated on a combined dataset from the Deepship and Ocean Networks Canada (ONC) benchmarks, providing a diverse set of real-world underwater scenarios. AquaSignal employs a U-Net architecture for denoising, a ResNet18 convolutional neural network for classifying known acoustic events, and an AutoEncoder-based model for unsupervised detection of novel or anomalous signals. To our knowledge, this is the first comprehensive study to apply and evaluate this combination of techniques on maritime vessel acoustic data. Experimental results show that AquaSignal improves signal clarity and task performance, achieving 71\% classification accuracy and 91\% accuracy in novelty detection. Despite slightly lower classification performance compared to some state-of-the-art models, differences in data partitioning strategies limit direct comparisons. Overall, AquaSignal demonstrates strong potential for real-time underwater acoustic monitoring in scientific, environmental, and maritime domains.

\end{abstract}

\end{frontmatter}

\section{Introduction}
Autonomous underwater acoustic detection and classification have emerged as key areas of research due to their broad applicability across critical domains such as maritime security, environmental monitoring, and natural resource exploration. Nations with extensive coastal frontiers, such as Australia and the United States, have significantly increased their investment in the development of advanced underwater surveillance systems aimed at detecting maritime vessels, including submarines and autonomous underwater vehicles (AUVs) \cite{smook05}. In parallel, oceanographers and climate scientists leverage passive and active acoustic sensing technologies to observe geophysical changes on the ocean floor and to study the behavior, distribution, and migration of marine species \cite{fmars2019}.

The offshore energy industry also relies heavily on acoustic surveys for applications such as seabed mapping, sub-seafloor imaging, and the structural integrity assessment of underwater infrastructure \cite{BUEGER2023}. In the domains of fisheries and aquaculture, real-time acoustic monitoring contributes to sustainable resource management by enabling the continuous observation of fish populations and farming conditions \cite{raq.12842}. In defense and security contexts, underwater acoustic signal detection and classification are vital components in tactical decision-making, providing situational awareness and threat identification in contested maritime environments \cite{bib9}. 

Furthermore, in geopolitically sensitive and environmentally significant maritime regions—such as the South China Sea, Arctic passageways, and the Gulf of Mexico—real-time underwater acoustic monitoring plays a critical role in maintaining national sovereignty, deterring illicit activities, and enforcing environmental regulations \cite{BUEGER2023, bib9, smook05}. These areas are increasingly subject to strategic competition, ecological vulnerability, and heightened maritime activity, necessitating robust surveillance systems that can operate autonomously and continuously under dynamic conditions.

Simultaneously, the rapid expansion of maritime traffic and subsea infrastructure, including underwater communication cables, oil and gas pipelines, and offshore renewable energy installations, has introduced new challenges in terms of operational safety, environmental risk, and asset protection. Effective acoustic monitoring is essential not only to detect and prevent accidental or intentional damage to these critical infrastructures but also to support predictive maintenance and reduce long-term operational costs \cite{BUEGER2023, raq.12842}.

Consequently, there is an urgent demand for robust, automated methods capable of not only accurately processing and interpreting the vast quantities of acoustic data generated by modern underwater sensor networks, but also serving as a promising solution for  denoising, classifying known underwater acoustic events, and detecting anomalous or novel signals \cite{bib12, Hee14}.

Recent advancements in underwater acoustic signal processing have increasingly incorporated Machine Learning (ML) methodologies to tackle complex tasks such as denoising, signal detection, and classification with greater accuracy and efficiency \cite{bib13, bib14}. In particular, Convolutional Neural Networks (CNNs) have gained prominence for their ability to extract hierarchical feature representations from raw acoustic data, thereby reducing input dimensionality, lowering computational overhead, and significantly improving classification performance in noisy and dynamic underwater environments.

Despite these advancements, several persistent challenges complicate the deployment of ML models in real-world marine settings. These include the presence of high levels of ambient and transient background noise, frequency-dependent attenuation of sound in water, a broad diversity of signal sources with overlapping spectral features, and the scarcity of labeled datasets, which limits supervised learning approaches and generalization across environments \cite{bib14}. Furthermore, conventional ML-based classification pipelines are typically trained on fixed, predefined classes and often lack mechanisms to handle novelty detection, such as identifying previously unseen or anomalous acoustic events.

To address these limitations, the present work introduces AquaSignal, a comprehensive ML-driven framework that integrates preprocessing, denoising, classification, and novelty detection within a single, unified pipeline. By combining existing state-of-the-art models and lightweight architectures, AquaSignal enhances the robustness and adaptability of acoustic analysis, offering improved performance across a range of underwater sensing scenarios, such as marine mammal monitoring, submarine volcanic activity detection, illegal fishing surveillance, and underwater vehicle localization. These diverse applications demonstrate AquaSignal’s capacity to operate effectively in both structured research settings and unpredictable, real-world marine environments \cite{bib14, Nils24, bib1, bib4, bib5, bib6, bib7}. Utilising publicly available underwater acoustic datasets, such as the Deepship datset \cite{bib1}, this paper proposes a novel four-stage pipeline comprising {\em Preprocessing}, {\em Denoising}, {\em Classification}, and {\em Detection} steps. In this approach, the audio signals collected using hydrophones are first segmented and reshaped during preprocessing, then they are filtered to eliminate background noise. Finally, they undergo classification and detection to determine whether they correspond to a known ship type or represent a new, previously unseen, detection. The main contributions of this work are: 
\begin{enumerate}
    \item The design, implementation, and empirical evaluation of AquaSignal, a novel signal processing pipeline that integrates preprocessing, denoising, classification, and novelty detection capabilities for underwater acoustic signals. The pipeline leverages state-of-the-art machine learning models to address challenges unique to the underwater domain, such as signal degradation, noise contamination, and limited annotated data.
    \item The demonstration of high accuracy in novelty detection using benchmark datasets, including Deepship \cite{bib1} and Ocean Networks Canada (ONC) \cite{Hee14} (described below), highlighting the pipeline’s ability to identify previously unseen acoustic patterns with strong generalization performance. This result underscores the framework's potential for real-world applications in anomaly detection and adaptive monitoring.
    \item The execution of a comprehensive ablation study to systematically assess the contribution of individual components within the AquaSignal architecture. This includes a comparative analysis of alternative classification and novelty detection models, providing insights into the models' performance trade-offs and advantages.
    \item The introduction and validation of a dedicated novelty detection algorithm specifically designed for underwater acoustic data. This component enhances the system’s robustness by enabling the identification of anomalous events beyond the scope of traditional supervised classification schemes.
\end{enumerate}

\section{Datasets and Sensors}

Recent research on underwater signal classification has used publicly available datasets to train various classification models \cite{bib14}. One such dataset, that will also be used in this work, is called Deepship \cite{bib1}, which contains 47.07 hours of audio recordings from four ship types: {\em Cargo}, {\em Tug}, {\em Passenger Ship}, and {\em Tanker}. As presented in \cite{bib1}, the Deepship dataset is composed of underwater recordings captured between 02 May 2016 to 04 October 2018 in strait of Georgia delta node using an icListen smart hydrophone. This sensor is a highly sensitive broadband digital hydrophone designed for ultra-quiet performance. It operates across a frequency range of 1 Hz to 12 kHz, with a dynamic range of 120 dB and a sensitivity of 170 dBV re. $\mu$ Pa. It requires a power supply of 12–24 Vdc, consuming 0.8 W. The housing is made of engineered plastic and titanium, allowing for operational depths of up to 200 meters and 3500 meters, respectively. This is an all-in-one compact device that integrates preamplifiers, filters, converters, and data transmission unit functionalities.

The data collection occurred over a period of approximately 29 months. Its deployment occurred in three phases: first, from May 2, 2016, to June 24, 2017, it was positioned at a depth of 141 meters at coordinates 49.080926666°N, 123.338713333°W; second, between June 24, 2017, and November 3, 2017, it was positioned at 49.08082191°N, 123.33923008°W at a depth of 147 meters; and third, from November 4, 2017, to October 4, 2018, it was placed at 49.080811°N, 123.3390596°W, at a depth of 144 meters. This data was collected by the Ocean Networks Canada (ONC) initiative \cite{Hee14}.

ONC is responsible for generating ‘big data’ in the form of high-resolution sensor measurements, video, and underwater sound recordings from the ocean \cite{Hee14}. Although the Deepship dataset does not contain samples specific to the `background' class, the general ONC database incorporates multiple recordings of underwater background noise, which were included in this work.

\section{Related Work} \label{sec:related_work}

Comprehensive surveys on deep learning methods and datasets for underwater acoustic classification can be found in \cite{bib14,Nils24}. Numerous strategies can be used to partition datasets into training and test sets. The most challenging one (which is used in this work) ensures that segments of an audio file appear exclusively in either the training or test set, never both. This prevents data leakage and forces the classification model to achieve high accuracy under stricter conditions. This section presents an overview of related research that applies this data partitioning method on the Deepship dataset.

Recent advancements in underwater acoustic signal classification have leveraged a range of deep learning models, each aiming to address challenges posed by complex signal characteristics such as high intra-class variability and noise interference. In a comparative study by \cite{bib4}, multiple classification models, including Random Forest, Support Vector Machines (SVM), Fully Connected Neural Networks (FCN), MobileNet-v3, and ResNet18 enhanced with Multi-Head Attention (ResNet18+MHA) were evaluated. The results reveal a progressive improvement in classification accuracy, from traditional machine learning approaches like Random Forest (66.71\%) and SVM (69.24\%) to deep learning models such as FCN (72.98\%) and MobileNet-v3 (70.45\%). However, the ResNet18+MHA model notably outperformed all others with an accuracy of 80.05\%. The integration of the multi-head attention mechanism appears to have played a pivotal role in this performance boost by enabling the model to suppress redundant features and better capture salient information from the acoustic signals.

Building on the theme of architectural specialisation, \cite{bib5} introduced the Convolution-based Mixture of Experts (CMoE) model, achieving a competitive accuracy of 79.62\%. Unlike conventional architectures, CMoE employs multiple expert layers as independent learners, guided by a routing mechanism that dynamically allocates input samples to the most suitable expert. This modular design enhances the model’s capacity to learn complex patterns in underwater acoustics, particularly those with high intra-class diversity, by facilitating specialisation and improving regularisation.

Efforts to improve classification performance and generalisability have led to the development of several novel architectures. The Adaptive Generalised Network (AGNet), for example, transforms fixed wavelet parameters into trainable components, improving its adaptability to diverse frequency characteristics in underwater signals \cite{bib6}. The Smooth-ResNet model introduced in \cite{bib7} incorporates a smoothness-inducing regularisation term, particularly leveraging simulated signals to enhance robustness. Additionally, the Separable Convolution AutoEncoder (SCAE) attempts to disentangle spatial and channel-wise features, which may offer improved interpretability and efficiency \cite{bib1}. These models achieved accuracies of 77.09\%, 78.25\%, and 77.53\% respectively, showing good performance but slightly trailing the leading methods in overall effectiveness.

In pursuit of even higher classification performance, \cite{bib21} proposed the Underwater Acoustic Learning Front-end (UALF) system, which integrates multi-perspective feature extraction with adaptive classifier configurations. This system reported the highest observed accuracy of 81.39\%, demonstrating the potential of flexible, end-to-end learning systems in adapting to diverse underwater acoustic scenarios. Similarly, \cite{bib22} explored a synchronous deep mutual learning framework that combines wave-based and time-frequency (T-F) representation models. This hybrid approach capitalised on the complementary strengths of the two representations, achieving a competitive accuracy of 79.50\%.

Collectively, these studies underscore the trajectory of model evolution in underwater acoustic classification, from conventional machine learning and standard deep learning architectures to specialised, adaptive, and ensemble-based methods. The consistent trend suggests that models incorporating domain-specific adaptations (such as multi-head attention, learnable front-ends, and expert-based routing) tend to outperform more generic architectures. The trade-off between architectural complexity and generalisation performance remains a key open issue, which the present work aims to solve.

All the classification methods discussed above use different attributes to partition and segment the audio files. These variations include segment sizes (e.g., 2s, 5s, or 30s), fixed or random training and test sets, and different class selections from the same dataset. These differences make impossible a fair comparison between the reported classification results. It is also worth noting that these results are based on a data partitioning strategy that prevents {\em leakage} between training and test sets, as previously mentioned \cite{bib14}. Other partitioning approaches have yielded higher accuracy values, such as the Sparse Representation-based Classifier (SRC) algorithm, reported in \cite{bib32}, which achieves an accuracy of 97.49\%. However, this performance is confounded by significant information leakage, as contiguous segments from the same audio recording are distributed across both the training and testing sets. This overlap undermines the validity of the reported results, inflating performance estimates and limiting the generalisability of the model to truly unseen data.

Finally, to the best of our knowledge, all research publicly available regarding the Deepship dataset is related to the task of identifying the different classes. Apart from the results reported in the present paper, there has been no other work addressing the challenge of novelty detection within this context.

The following section provides a detailed explanation of the machine learning models that constitute the core components of the AquaSignal architecture. These models were selected and integrated based on their demonstrated efficacy in addressing the unique challenges of underwater acoustic signal processing, including high ambient noise, signal variability, and limited labelled data. Each model is described in terms of its architectural design, training methodology, and specific role within the pipeline, encompassing the tasks of denoising, classification, and novelty detection.

\section{Machine Learning Models in AquaSignal}

\begin{figure*}[h]
\centering
\includegraphics[width=7.5in]{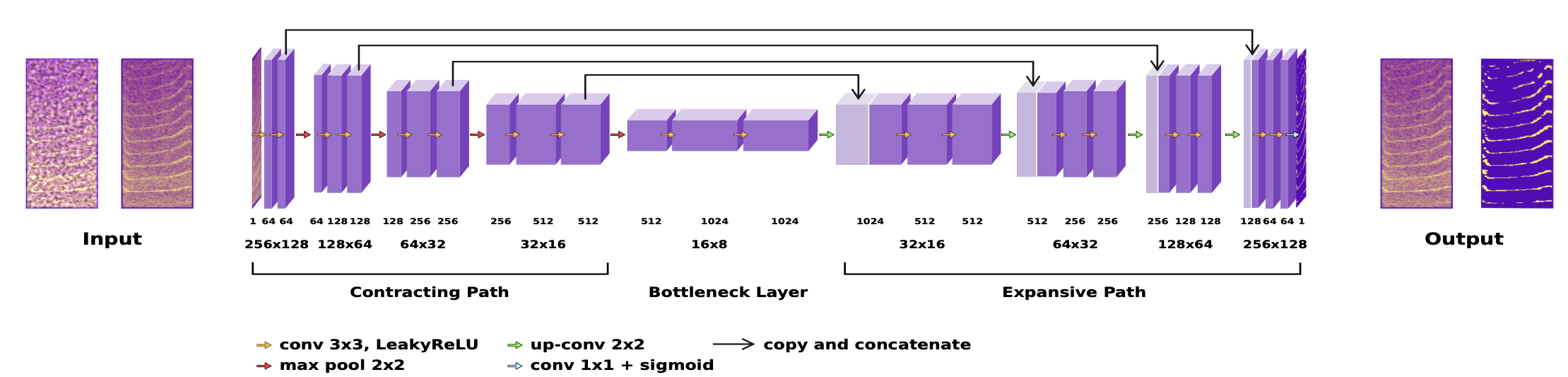}
\caption{ORCA-CLEAN – Deep denoising network architecture \cite{bib2}.}
\label{fig:ORCA_CLEAN}
\vspace{10pt}
\end{figure*}

Among the set of existing denoising methods for underwater acoustic signals \cite{gao2024underwater}, the ORCA-CLEAN algorithm proposed in \cite{bib2} has shown high accuracy with great generalisation capabilities. This is a fully-automated denoising method for bioacoustics based on a U-Net Deep Learning architecture \cite{gurrola2021}. This method does not require any denoised (clean) labelled ground-truth signals to be trained, and it consists of four stages: Preprocessing, Additive Noise Variants, Binary Mask Generation, and Deep Denoising Network. During the Preprocessing phase, the audio signal is first resampled to 32kHz, then the Short-time Fourier transform (STFT) is computed to generate a decibel-converted power-spectrogram, and lastly a random intensity, a pitch, and
a time augmentation are added to the samples, resulting in $2,049 \times T$ decibel-converted and augmented power-spectrograms, with $T=200$ being the number of analysed time frames. Next, various distributions of synthetic noise are utilised in order to corrupt the noisy original spectrograms, as well as real-world underwater
background noise. Before that, each spectrogram is compressed using linear frequency, resulting in a $256 \times 200$ spectrogram. Next is the Binary Mask Generation, where machine-generated binary masks are utilised, acting as a network attention mechanism. The original noisy spectrograms, together with the corresponding generated binary masks, enable different manifestations of potential denoised ground-truth samples \cite{bib2}. 
Moreover, ORCA-CLEAN framework is based on the U-Net architecture, where the network is trained on the $256 \times 200$ spectrograms. During training, the spectrograms input and output can be divided into two cases: (1) the input is the additive noise modified version of the spectrogram and the output is the original file, and (2) the original file is the input and the corresponding mask is the output \cite{bib2}. In both cases, the input spectrogram is noisier than the ground truth \cite{bib2}. Fig.\ref{fig:ORCA_CLEAN} illustrates this architecture.

\begin{figure}[h]
\centering
\includegraphics[width=3.5in]{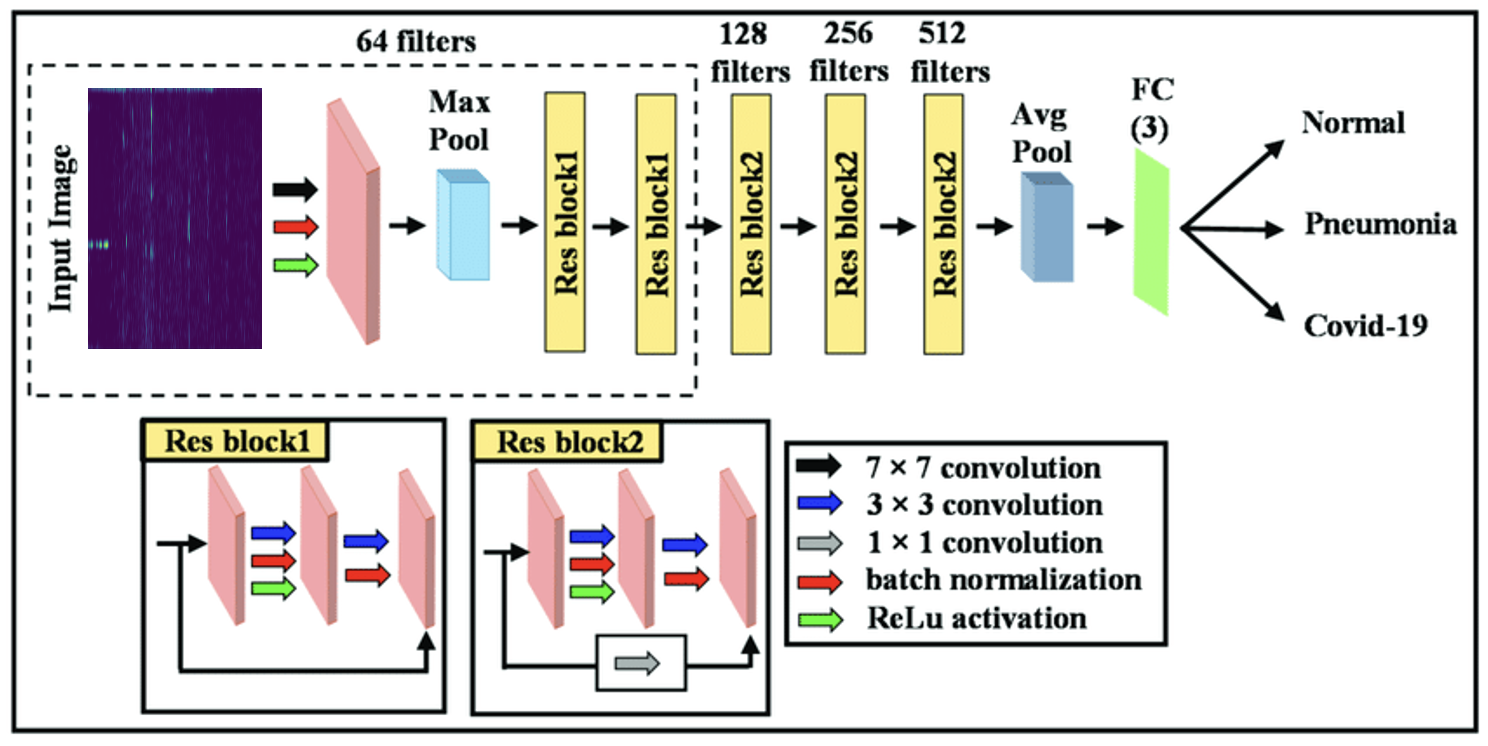}
\caption{ResNet18 Architecture \cite{bib19}.}
\label{fig:ResNet18}
\vspace{20pt}
\end{figure}

In the work reported in this paper, classification was executed by a Residual Network 18 (ResNet18) \cite{bib18}. As shown in Figure \ref{fig:ResNet18}, ResNet18 is a CNN consisting of blocks which are used to transfer residuals from previous layers to be used in the subsequent connected layers. This process enables the possibility to explore additional parts of the feature space which would have been missed in a conventional CNN. This model is defined by a 72-layer architecture with 18 deep layers, consisting of multiple shortcut connections and residual blocks. It was pre-trained on the ImageNet database and, in this work, it was fine-tuned with 3-channel spectrogram arrays. The ImageNet database is a large-scale, structured image database containing over a million images of one thousand different classes of visual objects. Using the pre-trained ResNet18 fine-tuned to a spectrogram dataset, transfer learning was applied, so the model could exploit the knowledge learned from a previous task to improve the generalisation in the current task.

In this research, we implemented a feedforward AutoEncoder (AE) architecture to perform unsupervised anomaly detection in the underwater acoustic data. The AutoEncoder is designed to learn a compact representation of background acoustic patterns, such that deviations from this representation can be flagged as anomalous. Structurally, the AE comprises three primary components: an encoder, a latent representation layer, and a decoder.

\begin{figure}[h]
\centering
\includegraphics[width=3.5in]{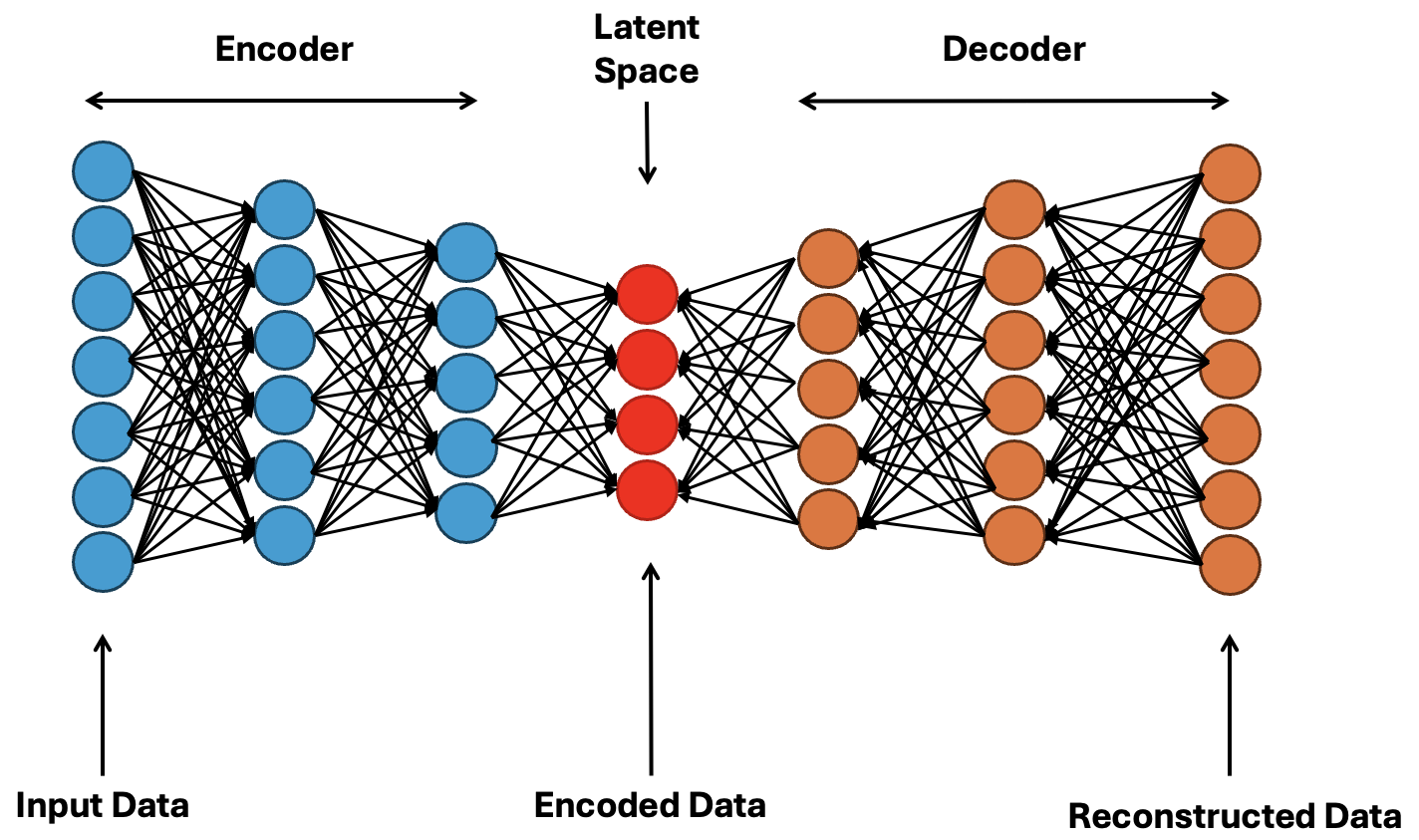}
\caption{AutoEncoder Architecture.}
\label{fig:VanillaAE}
\vspace{20pt}
\end{figure}

The encoder is composed of a sequence of fully connected dense layers that progressively compress the input vector, typically derived from flattened spectrogram representations, into a lower-dimensional embedding. This compression serves to extract the most informative and invariant features of the acoustic signal while discarding redundant or noisy information. The latent space, situated at the bottleneck of the architecture, encodes a compact set of high-level abstractions. These latent variables encapsulate the underlying statistical regularities present in the training data and form the basis for reconstruction. The decoder then maps this latent representation back to the original input dimensionality through a mirrored sequence of dense layers. Its objective is to reconstruct the original input with minimal information loss \cite{bib26}. As shown in Figure \ref{fig:VanillaAE}, the AutoEncoder architecture is symmetric with respect to the encoder and decoder paths. During training, the network minimises a reconstruction loss function, encouraging the latent space to model only the most salient features required for accurate reconstruction \cite{bib26}. Once trained on non-anomalous background data, the model can detect novel or anomalous events by measuring the reconstruction error. High errors indicate a mismatch between the learned data distribution and the input, thus signaling a potential anomaly.

This AutoEncoder module forms a critical component of the AquaSignal pipeline, as presented in the next section.

The next section presents a description of the technical parameters and implementation settings employed in the development of the proposed methods. This includes a thorough specification of training configurations, hyperparameter selections, and data preprocessing protocols, all of which are critical for ensuring the reproducibility, validity, and interpretability of the experimental results.

\section{Method}\label{sec:met}
\begin{figure}[h]
\centering
\includegraphics[width=3.6in]{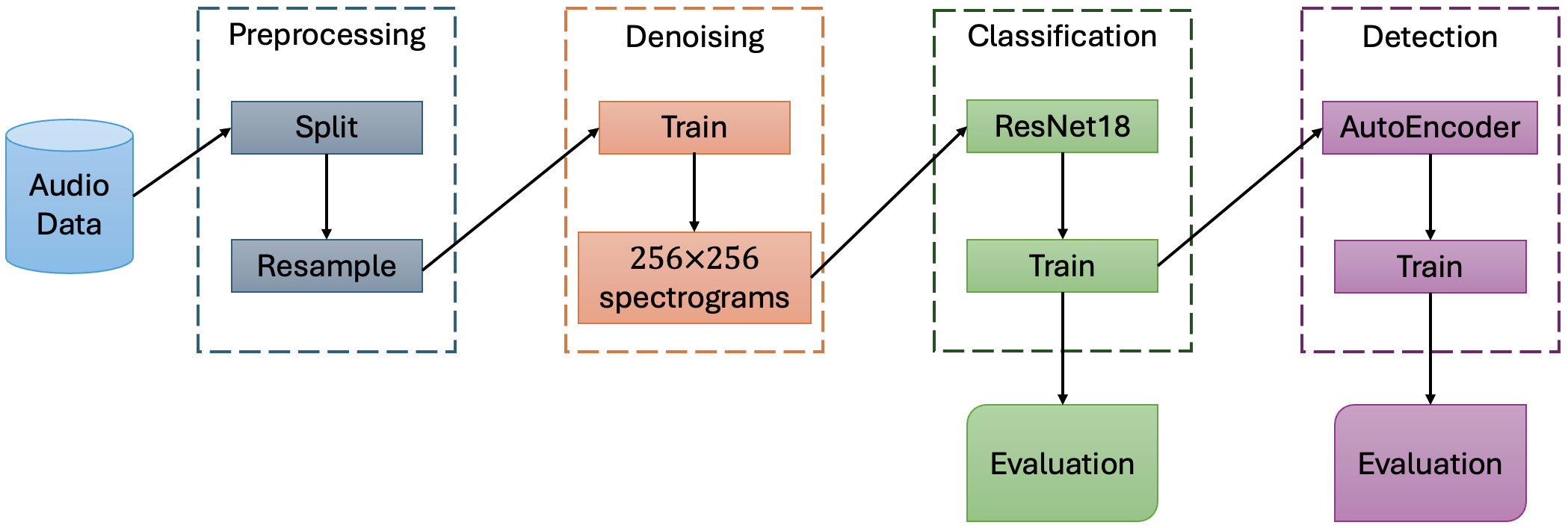}
\caption{Architecture framework of the proposed pipeline}
\label{fig:framework_arch}
\vspace{20pt}
\end{figure}

The AquaSignal pipeline proposed in this work is represented in Figure \ref{fig:framework_arch}. It consists of the following four stages: {\em preprocessing}, {\em denoising}, {\em classification}, and {\em anomaly detection}, as described below.

{\bf Preprocessing:}
Considering that the original DeepShip dataset does not include a background class representative of non-vessel or ambient maritime noise, additional background audio samples were sourced from the Ocean Networks Canada (ONC) database to supplement the dataset. These background recordings were incorporated during the data preprocessing phase to create a more realistic and comprehensive classification environment. Thus, the final curated dataset comprises five distinct classes: background, cargo, passenger ship, tanker, and tug.

As part of the data partitioning and preprocessing pipeline, the raw audio recordings were segmented into non-overlapping two-second clips to standardise the input length. These segments were then resampled to a consistent sampling frequency of 32 kHz to ensure uniformity across samples. This procedure yielded over 90,000 individual audio segments across all classes. To prevent data leakage and preserve the integrity of the model evaluation, it was ensured that the audio segments originating from the same original recording were not split between the training and test sets. This step was critical for guaranteeing that performance metrics accurately reflect the model's generalisation capability rather than its ability to memorise specific acoustic signatures.

For the novelty detection task, a single class was withheld during classifier training and reintroduced as contamination within the background data during testing. In this study, the `tug' class was designated for this purpose, representing 1\% of the samples within the background category. This setup simulates a realistic anomaly detection scenario, wherein rare but relevant events are embedded within ambient noise. To manage computational complexity and ensure a balanced class representation during training and evaluation, a stratified subsampling strategy was employed. Specifically, 5,000 audio samples were randomly selected from each of the five classes, resulting in a final working dataset consisting of 25,000 samples.

{\bf Denoising:}
As mentioned above, the denoising process in AquaSignal was executed using the ORCA-CLEAN model \cite{bib2}. The model was trained on the preprocessed dataset over the course of 100 epochs using the Adam optimiser with standard learning rate decay parameters. The objective was to adapt a denoising model capable of generalising across diverse underwater acoustic conditions without relying on clean ground-truth signals. To enhance the robustness of the model to realistic acoustic variability, each 2-second audio sample underwent a comprehensive augmentation pipeline prior to training. In addition to pitch shifting and amplitude scaling, which simulate Doppler effects and sensor sensitivity variation, time-stretching transformations were also applied within a ±10\% range to emulate variations in vessel speed and water column effects \cite{bib2, bib14}.
 
Moreover, Gaussian noise injection was used to model random thermal and electronic noise artifacts commonly found in hydrophone recordings, while a frequency masking technique inspired by SpecAugment—was employed to occlude specific frequency bands, replicating natural acoustic occlusion caused by environmental factors such as wave interference or overlapping signals from other marine entities. These transformations collectively aim to promote invariance in the learned features while preserving the essential spectral characteristics of ship-generated signals.

To ensure consistent scaling and suppress potential bias in the frequency response of the neural network, all power spectrograms were normalized on a per-frequency-bin basis, effectively standardizing the input distribution across different frequency channels. This normalisation not only facilitated more stable convergence during training but also contributed to improved generalisation in downstream classification and novelty detection tasks.

{\bf Classification:}
In this experimental setup, the `tug' class was treated as the contamination class. To simulate an out-of-distribution detection scenario, it was intentionally omitted from the training data and later introduced as 1\% of the total test samples. The dataset was then divided into training and testing subsets using a 90:10 split. For the classification task, a ResNet-18 model pretrained on ImageNet was chosen and fine-tuned on the training portion of the data.

A comprehensive hyperparameter tuning procedure was conducted to identify optimal training configurations, including the learning rate, batch size, and number of epochs. Based on the results of this parameter search, the model was trained for 50 epochs with a fixed learning rate of 0.001. The Cross-Entropy loss function was employed to quantify classification error, and the Adam optimisation algorithm was used to update model parameters during training. This setup was designed to ensure stable convergence while maintaining generalisation performance across both seen and unseen classes.

{\bf Detection:}
The AutoEncoder employed for novelty detection was trained in an unsupervised manner on a dataset predominantly composed of background class samples, which constituted 99\% of the data, with the remaining 1\% consisting of contamination samples introduced to simulate a realistic anomaly scenario. The AutoEncoder was designed to learn a compact representation of the background data distribution, enabling the identification of anomalous or out-of-distribution samples based on deviations in reconstruction error.

To establish a robust decision criterion for novelty detection, a reconstruction error threshold was defined using the 70th percentile of reconstruction error values computed exclusively from background samples. This threshold was selected to balance sensitivity and specificity in distinguishing novel inputs from those consistent with the training distribution.

Following an extensive hyperparameter optimisation and procedure—including exploration of learning rates, latent dimensionality, and architectural configurations, the final model was trained for 50 epochs. Training was conducted using the Adam optimisation algorithm and Binary Cross-Entropy as the loss function, which was appropriate for reconstructing input samples in the context of probabilistic pixel-wise or feature-wise reconstruction. This setup aimed to ensure stable learning and effective generalisation of the background data manifold while preserving sensitivity to anomalous patterns.

The ensuing section outlines a representative use case scenario that explicitly demonstrates the practical deployment of the AquaSignal pipeline within an operational, real-world context.

\section{Use Case Scenario: Port Security Monitoring}
To illustrate the practical utility and operational potential of the AquaSignal framework, consider a deployment scenario situated in a high-traffic international maritime hub, such as the Port of Singapore or the Port of Rotterdam. In this environment, a distributed network of hydrophones is installed across key monitoring zones to continuously capture real-time underwater acoustic data. The AquaSignal pipeline can be employed to autonomously process this incoming data stream through the following stages:

\begin{enumerate}
  \item \textbf{Preprocessing}: The continuous acoustic signal is segmented, normalised, and resampled into 2-second 32 kHz audio clips.
  \item \textbf{Denoising}: The processed audios are passed through the ORCA-CLEAN module, which is designed to suppress common forms of underwater acoustic clutter, including ambient wave noise, marine mammal vocalizations, while enhancing signal components associated with anthropogenic activity, such as ship propeller and engine signatures.
  \item \textbf{Classification}: The processed and denoised signals are analysed by the ResNet18 convolutional neural network, which classifies vessel types based on their acoustic signatures. Recognised categories include cargo ships, passenger ships, tankers, and tug boats.
  \item \textbf{Detection}: Finally, the AutoEncoder-based novelty detection module evaluates the input signals for deviations from known acoustic patterns. This mechanism flags previously unseen or anomalous sources, such as unauthorized mini-submarines or unidentified uncrewed underwater vehicles (UUVs), prompting human analysts for further inspection and verification.
\end{enumerate}

This deployment scenario demonstrates how AquaSignal can enable persistent, passive, and non-invasive underwater surveillance, effectively complementing traditional monitoring systems such as radar and Automatic Identification Systems (AIS). In doing so, the system reduces reliance on manual data annotation and enhances the capacity of maritime authorities and security units to detect and respond to potential threats or regulatory violations in real time \cite{smook05}.

The next section presents the test results obtained from the AquaSignal pipeline applied to the Deepship dataset.

\section{Results}\label{sec:res}
 
This section presents the results of testing the AquaSignal pipeline on the Deepship dataset.  Classification and detection were evaluated in terms of their Precision, Recall, and F1-Score \cite{bib14}.

\begin{figure}[h]
\centering
\setkeys{Gin}{width=\linewidth}
\begin{subfigure}{0.27\textwidth}
\includegraphics[width=1.6in]{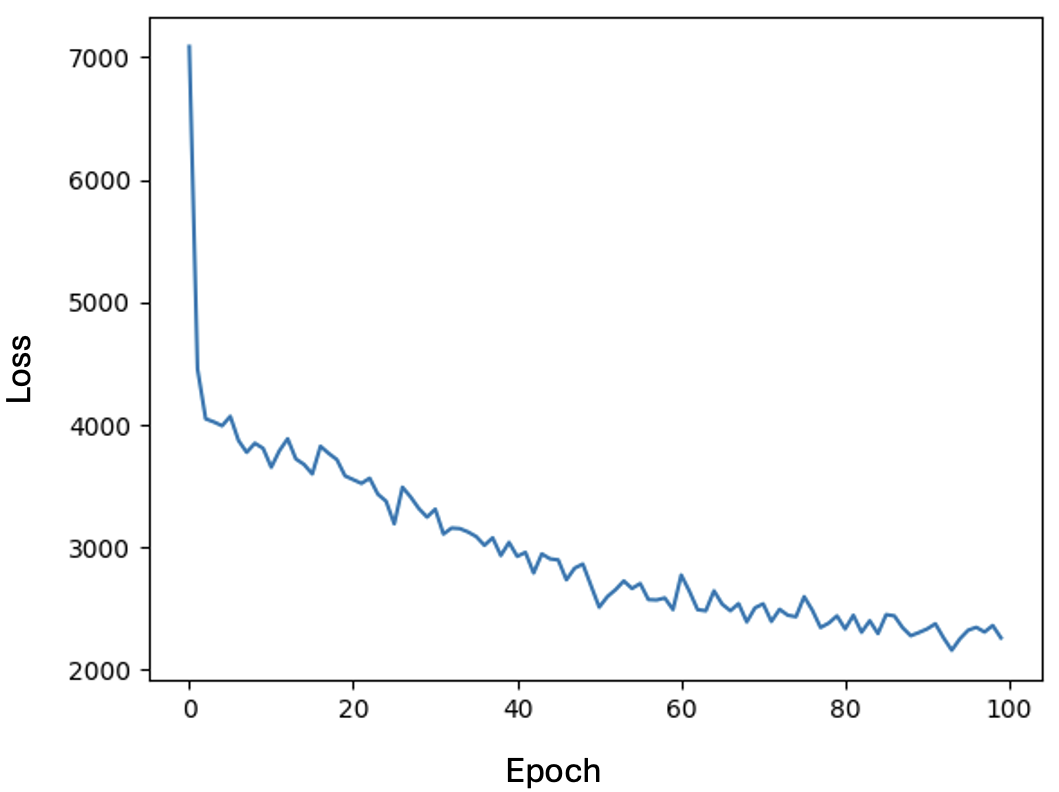}
\caption{Denoiser}
\label{fig:denoise_loss}
\end{subfigure}
\begin{subfigure}{0.25\textwidth}
\includegraphics[width=1.6in]{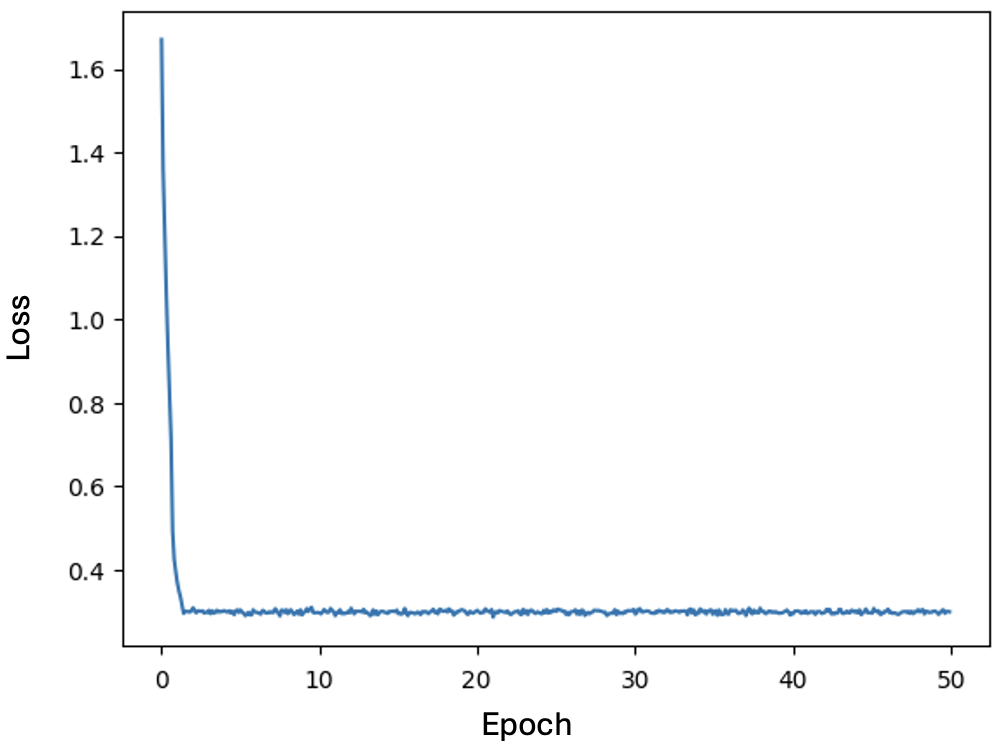}
\caption{Classifier}
\label{fig:loss_classifier}
\end{subfigure}
\begin{subfigure}{0.25\textwidth}
\vspace{10pt}
\includegraphics[width=1.6in]{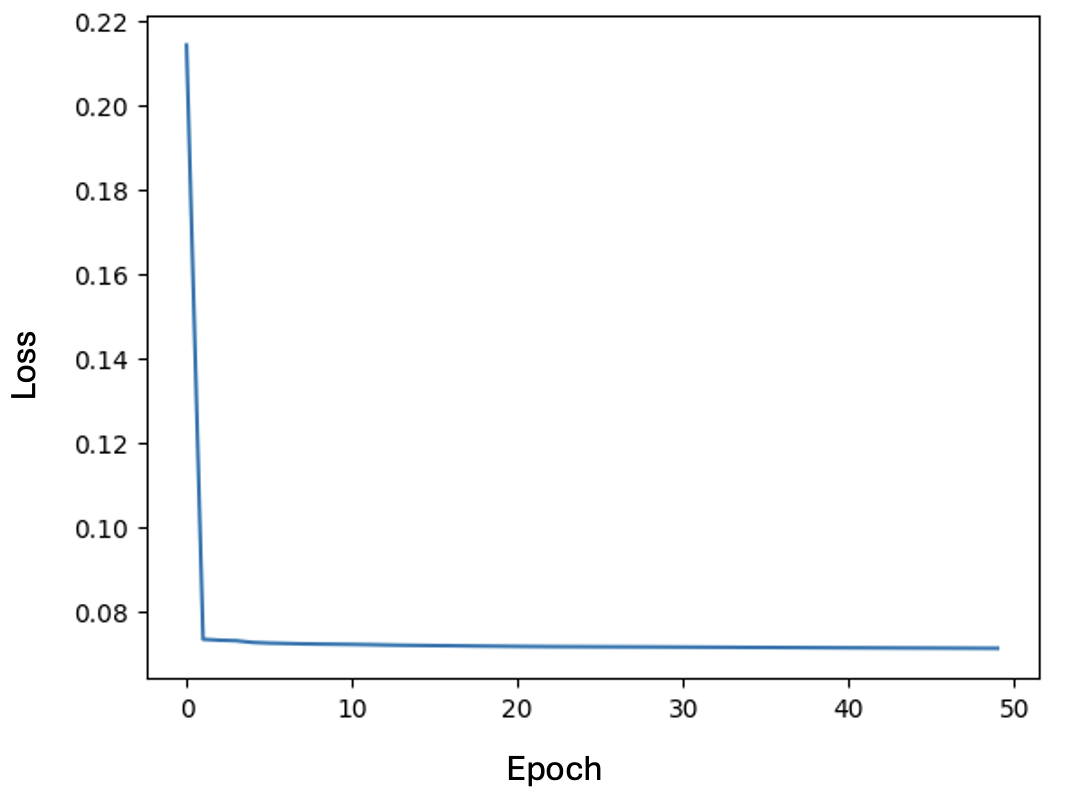}
\caption{Detection}
\label{fig:loss_detector}
\end{subfigure}
\vspace{10pt}
\caption{Loss plots}
\label{fig:csetup}
\end{figure}

{\bf Denoising:}
Figure \ref{fig:denoise_loss} presents the training loss curve of the denoising model over successive epochs, reflecting the model’s convergence behaviour and optimisation progress during training. This loss curve provides insight into the model’s ability to minimise reconstruction error when learning to distinguish structured signal content from background noise and stochastic interference.

\begin{figure}[h]
\vspace{20pt}
\centering
\includegraphics[width=3.2in]{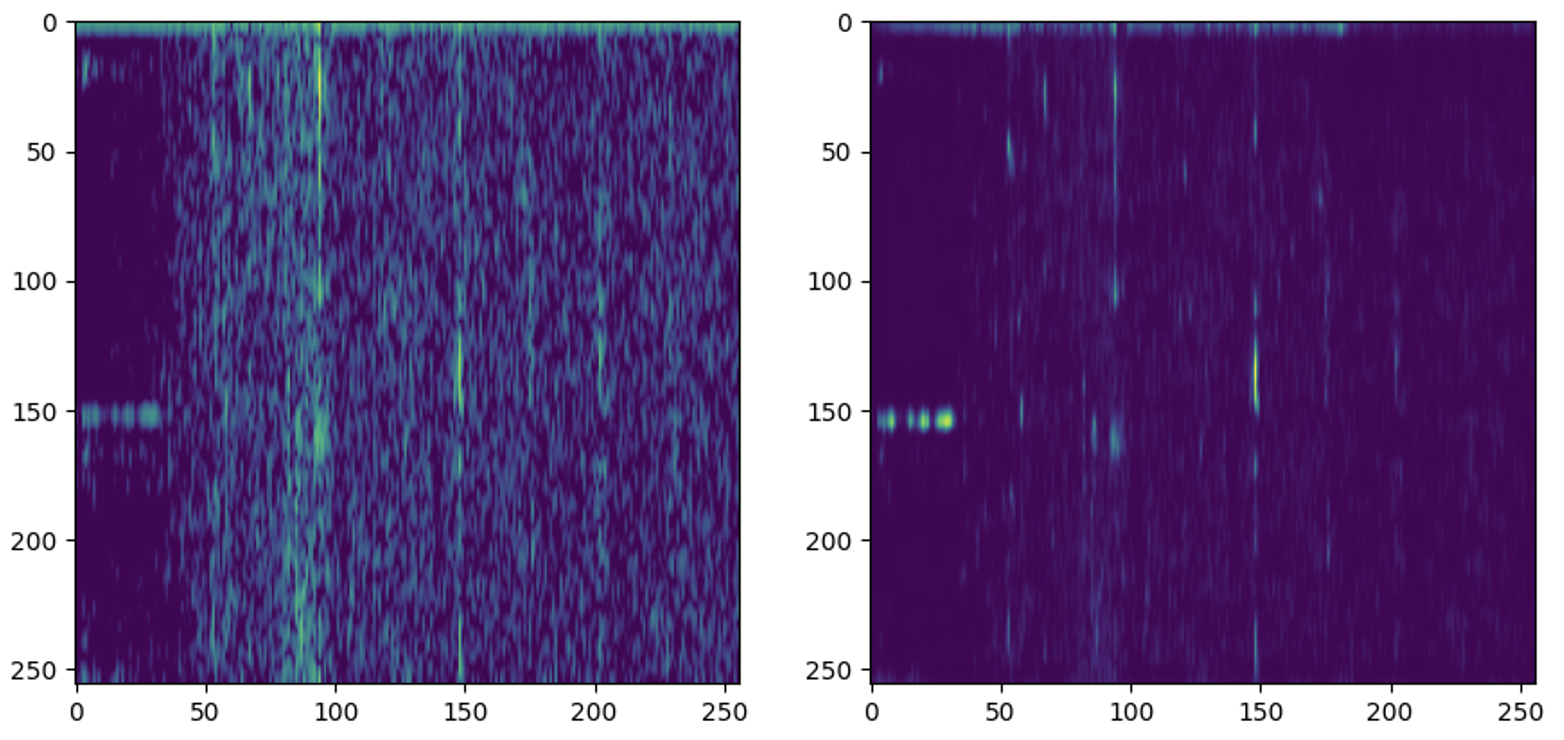}
\caption{Before (left) and after (right) denoising spectrogram}
\label{fig:denoise_before_after}
\vspace{20pt}
\end{figure}

Complementarily, Figure \ref{fig:denoise_before_after} depicts the time–frequency representation (spectrogram) of a sample from the background class, shown both prior to (left panel) and following (right panel) the application of the ORCA-CLEAN denoising algorithm. The pre-denoising spectrogram reveals a high degree of random interference and spectral clutter, characteristic of unfiltered marine acoustic environments. In contrast, the post-denoising spectrogram demonstrates a marked improvement in clarity, with a significant reduction in background noise and an enhanced delineation of salient frequency components.

Following denoising, distinct spectral features, potentially indicative of novel or anomalous acoustic events, become more readily observable. These results suggest that ORCA-CLEAN effectively suppresses unstructured noise while preserving signal components that may be critical for subsequent tasks in the pipeline.

{\bf Classification:}
Figure \ref{fig:loss_classifier} presents the training loss curve for the classification model, illustrating the convergence behaviour over 50 training epochs. The model demonstrates rapid convergence, achieving a plateau in the loss function within the first epochs. At convergence, the classifier attains a performance characterised by an accuracy of 0.713, precision of 0.719, recall of 0.725, and an F1-score of 0.710 on the test set. These metrics suggest a balanced performance, indicating that the classifier maintains a reasonable trade-off between false positives and false negatives across all classes.

\begin{figure}[h]
\centering
\includegraphics[width=3.2in]{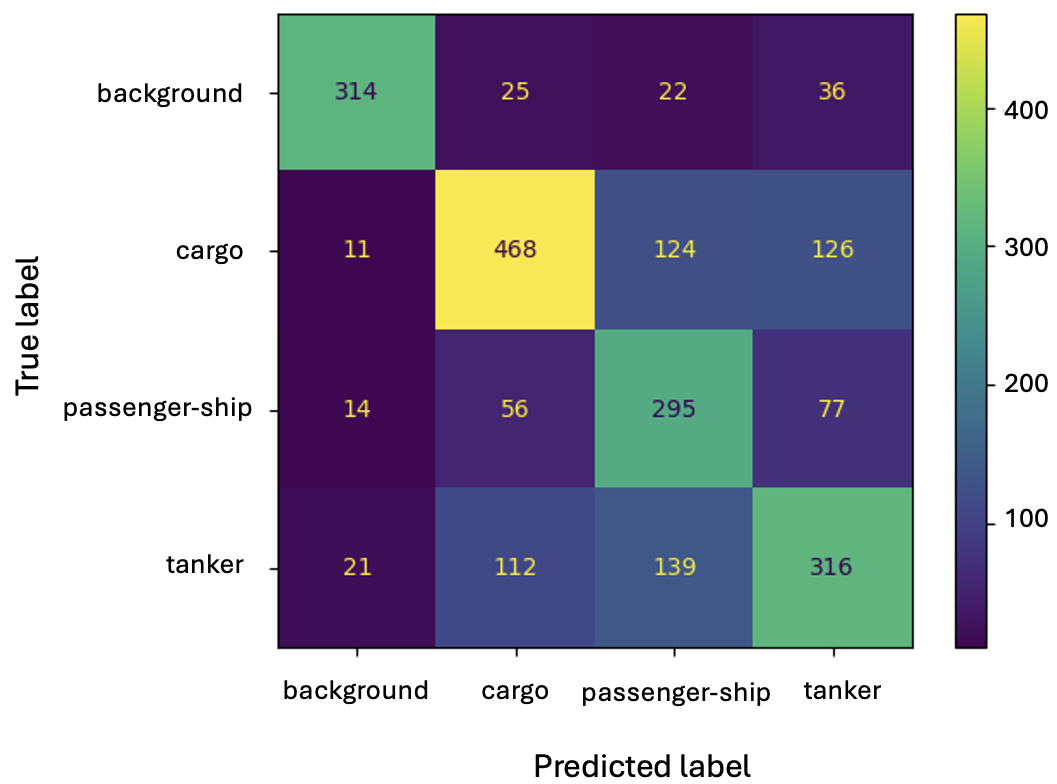}
\caption{Confusion matrix of the classification model}
\label{fig:classification_confusion_matrix}
\vspace{10pt}
\end{figure}

To further investigate the classification performance, the confusion matrix in Figure \ref{fig:classification_confusion_matrix} provides a detailed view of per-class predictive outcomes. The dominance of high values along the principal diagonal indicates that the majority of samples were correctly classified into their respective categories, reflecting the effectiveness of the learned discriminative features. However, the model exhibits a non-negligible error rate, particularly for the passenger-ship and tanker classes. These classes were frequently misclassified as either cargo or passenger ship, suggesting a representational overlap in the feature space or an underlying similarity in their acoustic signatures. Such confusion may stem from similarities in engine frequency components, vessel size, or propulsion patterns that challenge the classifier's ability to resolve fine-grained distinctions \cite{bib27, bib28, bib29}.

{\bf Novelty detection:}
Figure \ref{fig:loss_detector} illustrates the evolution of the training loss function for the detection model, highlighting the model’s convergence behaviour. The model demonstrates rapid convergence, reaching a near-optimal loss value within the first two epochs. This fast convergence suggests that the feature space, governing the binary detection task, discriminating between background noise and signal presence, is relatively well-structured and easily separable by the model's architecture.

At convergence, the model achieves a detection accuracy of 0.91. The discriminative performance of the model suggests that it effectively captures key spectral and temporal cues associated with active vessel or acoustic source presence.

To rigorously evaluate the performance of the proposed classification and novelty detection frameworks, an ablation study was conducted in which the developed models were benchmarked against several established baseline architectures. This comparative analysis aimed to quantify the efficacy of the proposed methods relative to state-of-the-art alternatives across a standardised evaluation protocol.

For the classification task, three widely recognized Convolutional Neural Network (CNN) architectures were implemented and fine-tuned for the acoustic signal classification problem: ResNet34 \cite{He_2016_CVPR}, DenseNet169 \cite{Huang_2017_CVPR}, and ResNet18 \cite{bib18}. Each model was trained and evaluated under identical conditions using the curated dataset described in Section \ref{sec:met}. As summarised in Table~\ref{table:model_comparison_classification}, the fine-tuned ResNet18 model, as proposed in this work, consistently achieved the highest performance across all key evaluation metrics, including Accuracy, Precision, Recall, and F1-Score. This outcome highlights the favorable balance of the model between representational capacity and generalisation, particularly in the context of relatively low-resource acoustic classification tasks.

In the context of novelty detection, additional comparisons were performed involving EfficientNet \cite{pmlr-v97-tan19a}, the Denoising Diffusion Anomaly Detection (DDAD) method \cite{mousakhan2023anomaly}, and a standard AutoEncoder (AE), which serves as the core detection module in the proposed AquaSignal framework. As presented in Table~\ref{table:model_comparison_detection}, the AE-based model implemented in this work outperformed the competing methods across all evaluation criteria. These included standard metrics such as Area Under the Receiver Operating Characteristic Curve (AUROC), Precision, Recall, and F1-Score, thereby demonstrating superior capability in distinguishing novel signals from background and known class distributions.

The results presented in the \ref{sec:related_work} section cannot be directly compared with those reported in the \ref{sec:res} section of this paper due to significant methodological differences in data partitioning strategies and dataset configurations. Specifically, the models in previous studies often relied on variable segment durations ranging from 2 to 30 seconds, random or fixed partitioning of the dataset, and differing class selections, all of which introduce inconsistencies that undermine comparability. Moreover, many of the reported results, even when using the same Deepship dataset, fail to guarantee strict separation of audio segments from the same original recording between training and testing sets. This introduces potential information leakage, which can artificially inflate classification performance. Additionally, AquaSignal uniquely incorporates a novelty detection component, absent in prior works, further differentiating its evaluation objectives and making direct metric-based comparisons methodologically inappropriate.


\begin{table}
\caption{Comparison of underwater signal classification models.}
\label{table:model_comparison_classification}
\centering
\begin{tabular*}{\columnwidth}{ccccc}
\hline
\hline
\textbf{Model} & \textbf{Accuracy} & \textbf{Precision} & \textbf{Recall} & \textbf{F1-Score}\\
\hline
ResNet34 & 66.9  & 66.5  & 67.8 & 66.8\rule{0pt}{10pt}\\
DenseNet169 & 69.24  & 66.71  & 66.71  & 66.71  \rule{0pt}{10pt}\\
\textbf{AquaSignal ResNet18} & \textbf{71.3} & \textbf{71.9}  & \textbf{72.5}  & \textbf{71} \rule{0pt}{10pt}\\
\hline
\hline
\end{tabular*}
\end{table}

\begin{table}
\caption{Comparison of underwater signal detection models.  }
\label{table:model_comparison_detection}
\centering
\begin{tabular*}{\columnwidth}{ccccc}
\hline
\hline
\textbf{Model} & \textbf{Accuracy} & \textbf{Precision} & \textbf{Recall} & \textbf{F1-Score}\\
\hline
EfficientNet & 29.9  & 51.3  & 51.5  & 29.8\rule{0pt}{10pt}\\
DDAD & 1.7  & 0.8  & 0.5  & 1.4  \rule{0pt}{10pt}\\
\textbf{AquaSignal AE} & \textbf{91.5}  & \textbf{87.8}  & \textbf{81.3}  & \textbf{84} \rule{0pt}{10pt}\\
\hline
\hline
\end{tabular*}
\end{table}


\section{Discussion and Concluding Remarks}

This paper presented AquaSignal, an achitecture for preprocessing, denoising, classifying, and detecting underwater acoustic signals. To the best of our knowledge, the proposed framework is the first to perform these tasks in a single pipeline. This work is also the first to execute novelty detection using the Deepship dataset. In this task, the results obtained show high accuracy in detecting novel signals in an unsupervised manner, even when using only a subset of the original dataset. The novelty detection task highlights AquaSignal's potential for monitoring environments where unseen vessels or unknown events must be detected. Additionally, ablation studies were performed showing that AquaSignal outperformed the state-of-the-art methods when trained using the same policy applied in this work. 

As shown by the evaluation (Section \ref{sec:res}), the AquaSignal architecture achieves strong performance across classification and novelty detection tasks, even with reduced data. Interestingly, as indicated in Tables \ref{table:model_comparison_classification} and \ref{table:model_comparison_detection} deeper networks such as ResNet34, DesNet169, EfficientNet, and DDAD underperformed, supporting the argument that data scarcity favors more efficient, shallower architectures. According to  \cite{bib24}, shallow networks are easier to optimise, train, and achieve high accuracy results, compared to deeper networks which face degradation problems, which implies that as the network depth increases, the accuracy becomes saturated and then degrades rapidly. 

Furthermore, there exists a trade-off in accuracy between denoised and noisy samples in the classification task. More specifically, when trained on the original noisy samples, the classifier returned an accuracy of 77.1\%, whereas when trained on the denoised data, the accuracy drops to 71.3\%. These findings indicate that, within the scope of the applications and datasets analysed in this study, the denoising process incurred a non-negligible loss of information. This outcome stands in contrast to prior reports in the literature, particularly those concerning the classification of biological signals, where denoising has been associated with improved or preserved informational integrity \cite{bib2}. A potential hypothesis for this observation could be that the classifier may have overfit to noise-related patterns that were consistent within classes but irrelevant to the real signal. In other words, certain noise artifacts, though technically not part of the biological signal, could have been correlated with specific classes due to how the data was collected or labelled. For example, if one class was collected in a noisier environment or with slightly different sensor settings, the noise profile itself becomes class-specific. Thereby, a classifier trained on this data might latch onto those noise patterns to help distinguish between classes. 

Despite the promising performance of the AquaSignal pipeline, several issues should be taken into account. First, while the denoising process reduces interference, it may also remove subtle features essential for class distinction, as reflected in the ~6\% drop in classification accuracy when denoised data is used. Second, the pipeline is currently tailored for ship-generated signals, extending it to biological or geophysical acoustic sources would require further domain adaptation \cite{bib2, bib1}.

In terms of future directions, future iterations of the detection task could benefit from incorporating temporal smoothing mechanisms or uncertainty estimation techniques, such as Monte Carlo dropout or Bayesian layers, to mitigate false alarms, particularly in borderline cases. Additionally, optimising the model on a class-balanced or cost-sensitive loss function may improve the handling of imbalanced detection scenarios, thereby enhancing robustness in operational deployments \cite{bib30, bib31}. Moreover, integrating multimodal inputs, such as video, could enable a richer understanding of underwater scenes. Additionally, online learning mechanisms can be incorporated to enable real-time adaptation as new classes or anomalies appear \cite{mousakhan2023anomaly, bib22}. Finally, methods like wavelet-based scattering transforms or complex-valued neural networks could be explored to capture finer-grained temporal structures \cite{bib13}.

\end{document}